\documentclass[sigconf]{acmart}
\usepackage[utf8]{inputenc}
\usepackage{booktabs}
\usepackage{enumitem}
\usepackage{multirow}
\usepackage{multicol}
\usepackage{xspace}
\usepackage{hyperref}
\usepackage{fontawesome5}

\newcommand{\method}{CultivAgents\xspace}

\AtBeginDocument{%
  }

\copyrightyear{2026}

\begin{document}

\title{CultivAgents: Cultivating Relationship-Centered Multi-Agent Systems for Personalized Gardening}
\subtitle{\vspace{0.5em}{\faIcon{seedling} Website: \url{https://hello-diana.github.io/CultivAgents/}\vspace{-0.5em}}}
\begin{teaserfigure}
  \includegraphics[width=\textwidth]{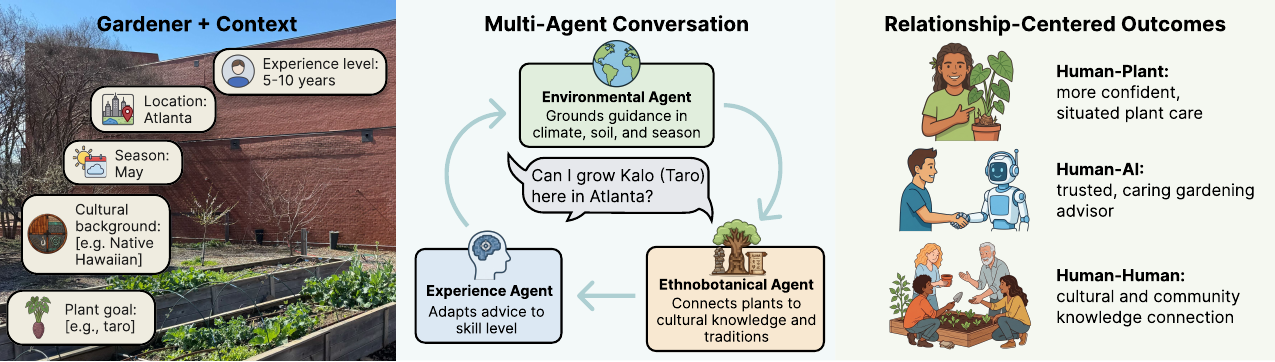}
  \caption{CultivAgents coordinates experience, environmental, and ethnobotanical agents to provide personalized gardening guidance that is actionable, locally grounded, and culturally meaningful.}
  \label{fig:hero}
\end{teaserfigure}



\author{Yiyang Wang}
\orcid{0000-0001-5778-3251}
\affiliation{%
   \institution{Georgia Institute of Technology}
   \city{Atlanta}
   \state{Georgia}
   \country{USA}}
\email{ywang3420@gatech.edu}

\author{Moeiini Reilly}
\orcid{0009-0004-8723-7146}
 \affiliation{%
 \institution{Georgia Institute of Technology}
 \city{Atlanta}
 \state{Georgia}
 \country{USA}}
 \email{moi@gatech.edu}

 \author{Britney Johnson}
 \orcid{0000-0002-3681-9742}
 \affiliation{%
  \institution{Georgia Institute of Technology}
   \city{Atlanta}
   \state{Georgia}
   \country{USA}}
 \email{bjohnson324@gatech.edu}

 \author{Kefei Yan}
 \orcid{0009-0002-2274-0467}
 \affiliation{%
  \institution{Georgia Institute of Technology}
  \city{Atlanta}
  \state{Georgia}
  \country{USA}}
  \email{kyan66@gatech.edu}

\author{Alex Cabral}
\orcid{0000-0002-6882-6701}
\affiliation{%
  \institution{Massachusetts Institute of Technology}
  \city{Cambridge}
  \state{Massachusetts}
  \country{USA}
}
\email{alcabral@mit.edu}

\author{Josiah Hester}
\orcid{0000-0002-1680-085X}
\affiliation{%
\institution{Georgia Institute of Technology}
\city{Atlanta}
\state{Georgia}
\country{USA}}
\email{josiah@gatech.edu}



\renewcommand{\shortauthors}{Wang et al.}

\begin{abstract}

Gardening is critical to support well-being, cultural continuity, and food autonomy, yet existing digital tools often provide generic advice that overlooks gardeners' skills, local ecologies, seasons, and cultural contexts. We introduce CultivAgents, a relationship-centered multi-agent system for personalized, socio-culturally grounded gardening support.
Grounded in \emph{ethics of care}, CultivAgents coordinates multiple specialized agents: an Experience Agent that adapts guidance to users' skill levels, an Environmental Agent that grounds advice in local and seasonal conditions, and an Ethnobotanical Agent that connects plants to cultural knowledge and histories. We evaluated CultivAgents through a three-phase mixed-methods study with domain experts ($n=3$), HCI researchers ($n=7$), and community gardeners ($n=5$), analyzing expert feedback, pre/post surveys, and participatory design activities.
Results suggest that CultivAgents helped gardeners translate interest into situated action: community gardeners reported increased confidence (3.00 to 3.60), motivation (4.00 to 4.40), and trust in acting on AI advice (3.20 to 4.00). Participants valued hyperlocal ecological guidance and complementary agent perspectives, while also identifying limits in cultural specificity, ecological grounding, and agent coordination. The work advances relationship-centered AI, offering design implications for multi-agent systems that support food sovereignty, community resilience, and cultural preservation.
\end{abstract}




\maketitle

\section{Introduction}
Gardening is a fundamental practice for supporting well-being, self-sufficiency, community resilience, and food sovereignty. By helping people grow food, care for living environments, and maintain connections to place, culture, and community, gardening goes beyond an individual leisure activity and 
serves as a source of accomplishment, relaxation, and control~\cite{pantiru2024umbrella, goralnik2023growingcommunity}. 
 In this sense, community and home gardens can contribute to local food access, green space, and community capacity, especially in urban and resource-constrained contexts. 
However, sustaining gardening practice is difficult, especially for new or geographically displaced growers. Novice gardeners have limited experience, and roughly half of new food gardeners quit after their first season~\cite{goralnik2023growingcommunity}. Geographically displaced growers may face unfamiliar climates, soil and pest conditions, or reduced access to the plants, materials, and cultural knowledge that shaped how they learned to grow food. 
Thus, supporting personal food growing requires more than general plant-care instructions, but locally grounded, culturally meaningful guidance that evolves with gardeners' expertise, resources, and goals. 

Recent digital gardening tools and plant-care chatbots powered by large language models (LLMs) make gardening advice more portable and accessible but face two limitations. First, \emph{gardening support is expertise-intensive}. Actionable guidance must account for plant biology, local climate, soil conditions, seasonality, pests, available materials, and culturally specific plant knowledge~\cite{otoadese2025growingroots,goralnik2023growingcommunity}. Ungrounded advice can lead to wasted effort, failed crops, inappropriate chemical use, or unsafe plant-care decisions~\cite{tzachor2023nature}. Second, \emph{gardening is relational and situated}. Useful support must balance what is environmentally feasible, appropriate to the gardener's experience level, and meaningful within their cultural or community context. Most LLM-based gardening systems still operate as standalone general-purpose chatbots, which collapse these dimensions into generic advice and overlook the social, ecological, and cultural relationships that shape gardening practice~\cite{owiti-kipkebut-2025-enhancing,nguyen2025my}. 

Multi-agent systems can address this gap by making different forms of expertise explicit rather than forcing them into a single generalized response. Building on this direction, CultivAgents explores how multi-agent LLM systems can provide personalized gardening support by coordinating ecological expertise, user experience, and socio-cultural knowledge within situated conversation.

In this paper, we present CultivAgents, a multi-agent LLM system that coordinates complementary perspectives for personalized gardening support. CultivAgents combines three specialized agents: an Experience Agent, which adapts guidance to the gardener's skill level and confidence; an Environmental Agent, which grounds recommendations in local climate, seasonal, and ecological conditions; and an Ethnobotanical Agent, which surfaces cultural narratives, traditional knowledge, and relationships between plants, place, and community. By coordinating these agents, CultivAgents provides personalized checklists, local insights, and cultural stories that move LLM gardening assistance beyond transactional Q\&A toward a compassionate companion for sustaining engagement and strengthening gardeners' connections to the environment.


We evaluated CultivAgents via a three-phase mixed-methods approach with 3 domain experts, 7 human-computer interaction (HCI) researchers, and 5 community gardeners. 
Across these phases, we analyzed expert feedback, surveys, and workshop activities to examine how the coordination of specialized agents shaped gardeners' motivation, confidence, trust, and perceived relationships with plants, place, culture, and community.
We found that CultivAgents significantly bolstered user confidence (mean increase from 3.00 to 3.60), motivation (mean increase from 4.00 to 4.40), and trust to act (mean increase from 3.20 to 4.00) by providing ``hyperlocal" ecological anchoring. Our analysis revealed that the system’s value lies in its ability to move the plant from a digital abstraction to a situated living entity through geographic and cultural specificities. Furthermore, while initially perceived as a ``trusted advisor," the system facilitated unexpected "heritage discovery," allowing users to connect gardening decisions with cultural knowledge and explore ancestral knowledge frameworks as a lens for personalized food production. 

This work contributes: (1) the design and implementation of CultivAgents, a multi-agent LLM system for personalized gardening support; (2) empirical insights into how specialized agents can support gardeners' confidence, motivation, and relationships with plants, place, culture, and community; and (3) design implications for relationship-centered multi-agent systems that support situated agency, cultural preservation, and food-sovereignty-oriented personal food growing.

\section{Related Work}

\subsection{LLMs and Multi-Agent Systems} 
General-purpose LLMs struggle to provide context-aware and actionable guidance~\cite{tzachor2023nature}. 
Prior systems improve reliability by incorporating retrieval augmented generation (RAG)~\cite{singh2024farmer,fanuel2025agriregion}, continuous adaptation~\cite{balaguer2024rag}, domain-specific models~\cite{yang2024pllama,awais2025agrogpt}, and modular tool-calling~\cite{yang2024shizishangpt}. 
While these systems improve factual grounding for agricultural extension, crop management, and plant diagnosis, they largely focus on commercial agriculture and frame users as information seekers, rather than gardeners engaged in sustained ecological, cultural, and emotional care~\cite{krzyzewska2025sustainability}. 
Meanwhile, home-gardening assistants usually use single-agent designs that collapse plant diagnosis, environmental reasoning, user memory, and culturally sensitive encouragement into one generic voice~\cite{tomar2024ecosage}. 

Multi-agent systems (MAS) offer a natural solution by distributing expertise across role-specialized agents~\cite{guo2024multiagentsurvey,wang2026mascot}. 
Generative Agents~\cite{park2023generativeagents}, AutoGen~\cite{wu2024autogen}, MetaGPT~\cite{hong2024metagpt}, and CAMEL~\cite{li2023camel} demonstrate persistent personas, flexible multi-agent dialogue, and workflow coordination. 
CompanionCast~\cite{wang2025companioncast} 
shows that role-specialized agents can improve perceived social presence in social co-viewing. 
However, most MASs target task completion, workflow execution, or simulation rather than long-term situated care~\cite{murad2026agenticai}. 
\method addresses this gap~\cite{li2024goodit} by coordinating experience, environmental, and ethnobotanical agents for personalized gardening support grounded in local ecology, user expertise, and cultural context.

\vspace{-4mm}
\subsection{Food Sovereignty and Gardening Praxis}
Gardening is culturally meaningful, ecologically situated, and relational. For immigrant and displaced communities, it can preserve cultural foodways, rebuild social ties, and transmit knowledge across generations~\cite{otoadese2025growingroots,onyango2025scoping}. However, displacement can interrupt this praxis: culturally significant crops may be unavailable, familiar growing knowledge may not transfer to new soils and climates, and gardeners may struggle with unfamiliar pests, seasons, and local conditions~\cite{goralnik2023growingcommunity}. Supporting displaced and culturally diverse gardeners is therefore connected to food sovereignty, cultural continuity, and public health equity~\cite{wright2021culturalfood,gangamma2024refugees,pantiru2024umbrella}.

Existing food and agricultural technologies often overlook cultural practices, social relations, and lived conditions~\cite{talhouk2022refugee,doggett2024migrant}, while HCI agriculture research has tended to prioritize commercial farming over urban and community gardening~\cite{doggett2023hciagriculture}. Although human--plant interaction and more-than-human design expand how HCI conceptualizes plants as living participants in care~\cite{chang2022patterns,loh2024morethanhuman}, these insights have rarely shaped AI systems that combine local ecological advice with culturally grounded gardening support.

The design of \method draws on the \emph{ethics of care}, which frames care as situated, relational, and responsive to vulnerability. Care ethics connects gardening to responsibility, reciprocity, and learned attentiveness~\cite{sovova2021care}, while care-centered AI and HCI emphasize attentiveness to situated needs, responsible intervention, competence, and responsiveness to lived experience~\cite{villegas2023moraldistance,umbrello2021vsd,henriques2025feminist}. Related calls for pluralistic and Indigenous-informed AI further caution against decontextualized assumptions about knowledge and expertise~\cite{lewis2024abundantintelligences,perera2025indigenous}. Building on these perspectives, \method explores how an MAS can ground gardening advice across experiential, environmental, and ethnobotanical perspectives. 

\section{Methods}

CultivAgents is a multi-agent system that provides personalized, socio-culturally grounded gardening guidance through a conversational web interface. Inspired by previous role-specialized multi-agent architectures~\cite{wang2025companioncast}, the system coordinates multiple LLM agents through selector-based group conversations. 

\subsection{System Implementation}

\subsubsection{User Onboarding and Profile Injection.}
\begin{figure}[h]
\vspace{-2mm}
\includegraphics[width=8cm]{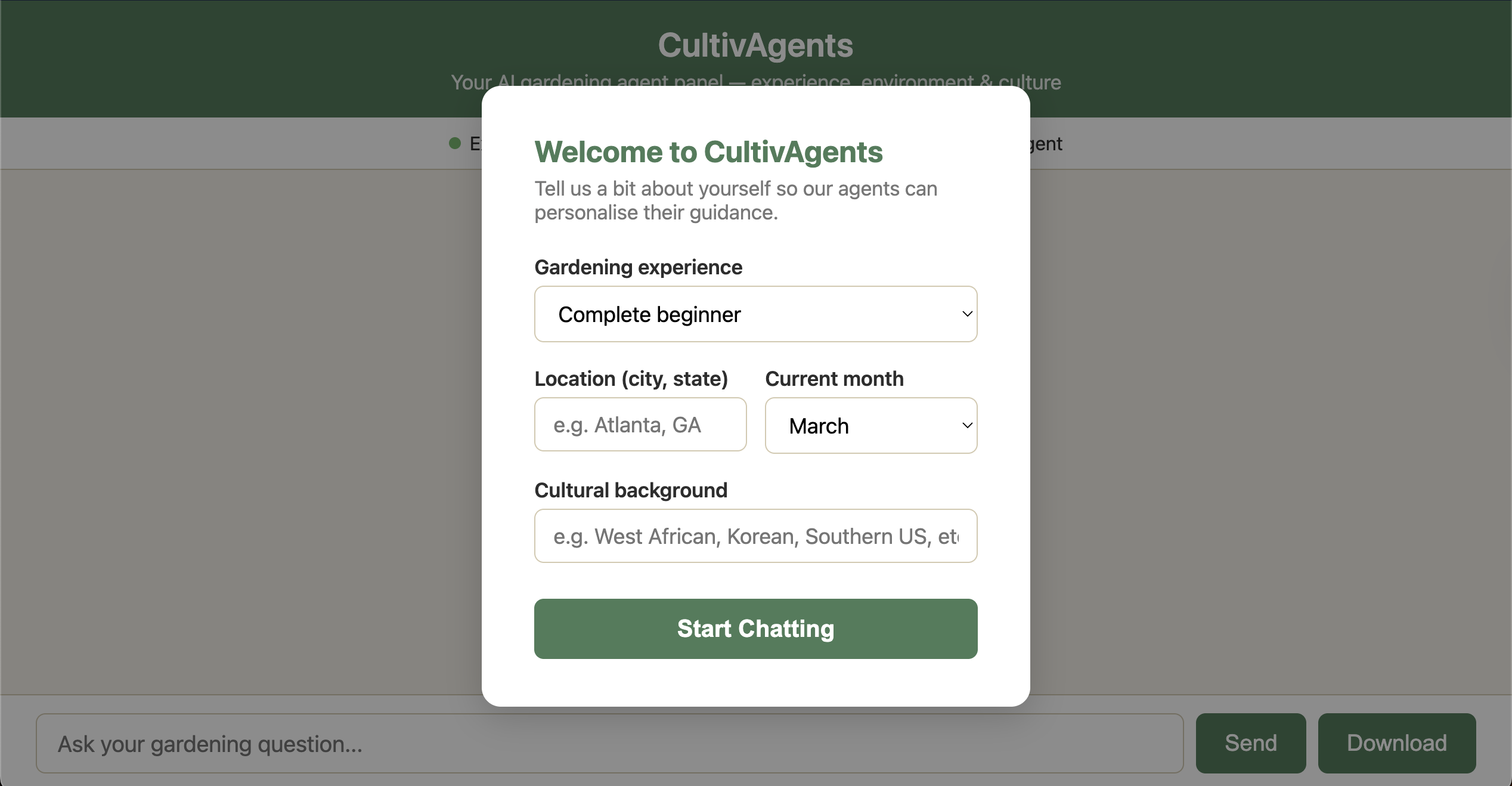}
\caption{Onboarding dialog collecting gardening experience level, location, current month, and cultural background.}
\label{fig:onboarding}
\vspace{-2mm}
\end{figure}

At the session beginning, a modal dialog (Figure~\ref{fig:onboarding}) collects a user profile:
\begin{equation}
  \mathbf{u} = \langle \, e, \; \ell, \; m, \; c \, \rangle,
\end{equation}
where $e$ is gardening experience level, $\ell$ geographic location, $m$ current month, and $c$ cultural background. Each agent $a_i \in \mathcal{A}$ 
has a base system prompt $s_i$ that defines its persona and guidelines. 
At session initialization, the effective prompt becomes:
\begin{equation}
  s_i^{*} = g_i(s_i) \;\|\; f(\mathbf{u}),
\end{equation}
where $f(\cdot)$ serializes the user profile into structured natural-language context. $\|$ denotes string concatenation. $g_i$ resolves agent-specific template variables, such as injecting the current date for the Environmental Agent. 
The resulting prompt $s_i^{*}$ personalizes $a_i$ with the same user profile while preserving its specialized role. 

\subsubsection{Agent Design.}
\method is designed as an extensible multi-agent framework. 
In this study, we instantiate three agents that cover complementary dimensions of the gardening experience:
\begin{itemize}[leftmargin=1em]
  \item \textbf{Experience-Level Agent} ($a_{\text{exp}}$) adapts complexity and tone to the gardener's skill level, ranging from step-by-step guidance for novices to concise, technical advice on cultivars, soil amendments, and pest management for experienced gardeners.

  \item \textbf{Environmental Context Agent} ($a_{\text{env}}$) provides location-specific, seasonal guidance, referencing local soil conditions, frost dates, and extension resources. 

  \item \textbf{Cultural and Ethnobotanical Agent} ($a_{\text{eth}}$) shares historical and medicinal plant knowledge across cultures, distinguishing traditional uses from scientifically validated procedures.
\end{itemize}

\subsubsection{Agent Coordination.}
At each turn $t$ within a response round, an LLM-based selector $\mathcal{M}_{\text{sel}}$ chooses the next speaker 
based on the conversation history $H_t = [h_1, h_2, \ldots, h_{t-1}]$ and the eligible agent set $\mathcal{A}_t \subseteq \mathcal{A}$. To encourage complementary perspectives, the eligible set excludes the immediately previous speaker $a_{t-1}$: $\mathcal{A}_t = \mathcal{A} \setminus \{a_{t-1} \}$ for $t \ge 2$. 
The selected speaker is then: 
\begin{equation}
  a_t = \mathcal{M}_{\text{sel}}\!\left(p_{\text{sel}},\; H_t,\; \mathcal{A}_t\right),
\end{equation}
where $a_t$ is the selected speaker, and $p_{\text{sel}}$ is 
the selector
prompt that routes questions according to agent roles: technical questions to $a_{\text{exp}}$, seasonal or location-specific queries to $a_{\text{env}}$, and cultural questions to $a_{\text{eth}}$. 
Each round terminates after $k{=}3$ agent messages. 
We then reset the round-level selector history ($H \leftarrow \emptyset$) to bound context growth while preserving conversational continuity.


\subsubsection{Communication and Frontend.}
\begin{figure}[h]
\includegraphics[width=8cm]{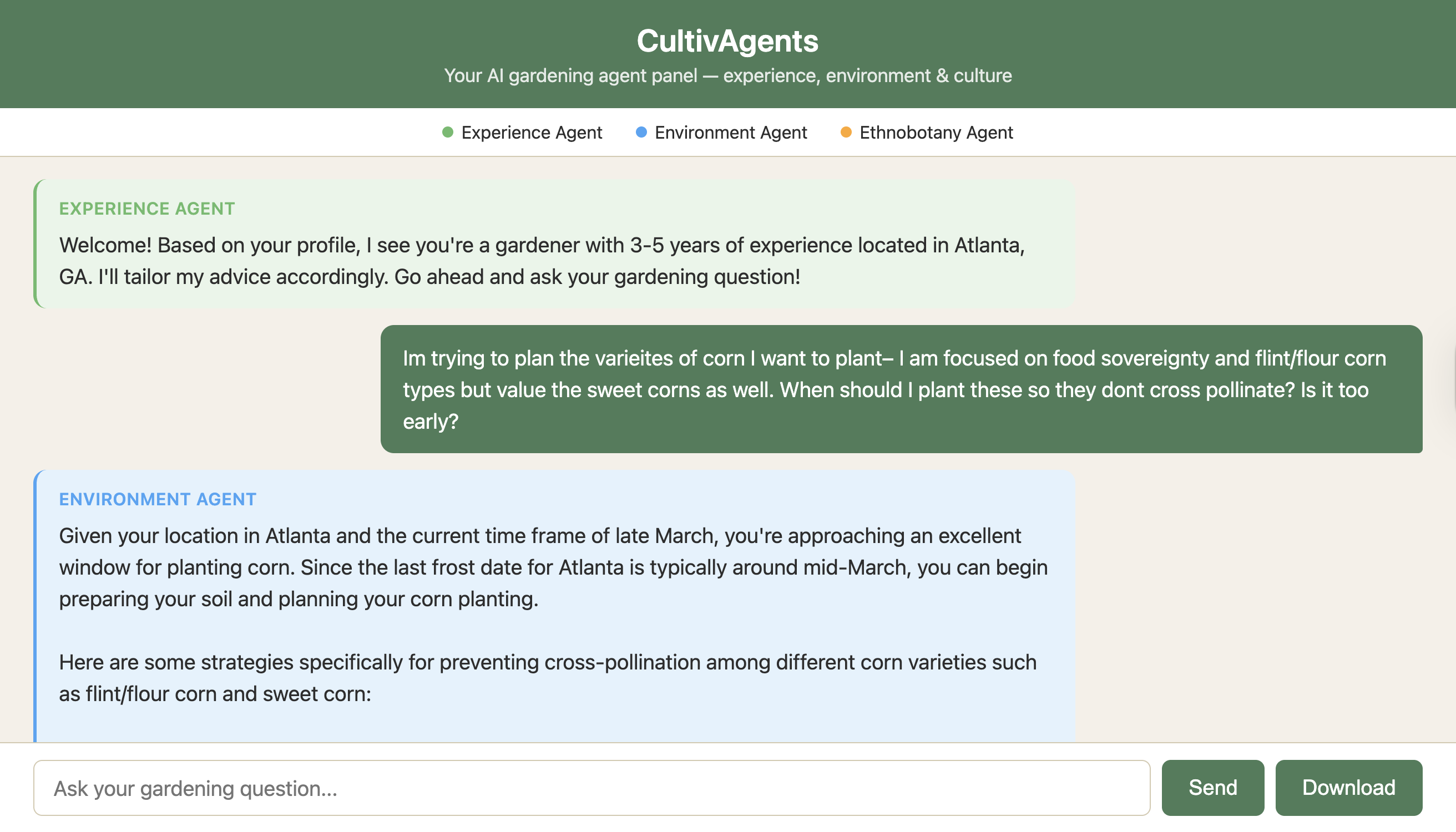}
\vspace{-2mm}
\caption{Chat with color-coded agent messages: green (Experience), blue (Environment), orange (Ethnobotany).}
\label{fig:system_mockup}
\end{figure}

\method uses a single-page frontend and FastAPI/WebSocket backend to stream AutoGen AgentChat~\cite{wu2024autogen} responses into a color-coded, markdown-supported chat interface (Figure~\ref{fig:system_mockup}). Users can also export conversations as text. The system is containerized via Docker and deployed on Render\footnote{\url{https://cultivagents.onrender.com/}}.

\subsection{Study Design}

We conducted a 3-phase mixed-methods study to iteratively evaluate and refine CultivAgents. 

\textbf{Phase I: Expert Feedback.} 
The first phase gathered expert feedback on whether CultivAgents could support culturally situated and locally specific gardening concerns. This phase focused on the system's cultural knowledge, ecological specificity, and relevance to food-growing practices across different regional contexts.

\textbf{Phase II: Formative Evaluation with HCI Researchers.} 
The second phase evaluated the usability, interaction flow, and clarity of CultivAgents' multi-agent design. This phase helped identify early interaction issues before deploying the system with community gardeners.

\textbf{Phase III: Participatory Evaluation with Community Gardeners.} 
The third phase evaluated CultivAgents with community gardeners and used participatory design to explore future relationship-centered agent personas and support needs. This phase focused on how gardeners perceived the system in relation to their own gardening practices, plants, cultural knowledge, and community.

\subsection{Participants and Recruitment}

Study procedures were approved by the Georgia Tech Institutional Review Board (IRB). Participants were recruited through purposive sampling based on their relevance to each study phase. All participants reviewed an informed consent form before participation.

\textbf{Phase I.} We recruited values-aligned domain experts ($n = 3$) operating as technological innovators and cultural stewards in Hawai`i and Georgia. We recruited these individuals due to their socio-technical capacities as agricultural technology designers and implementers working within sovereignty and education movements.

\textbf{Phase II.} We recruited HCI researchers ($n=7$) at Georgia Tech who could evaluate CultivAgents both as potential users and as researchers familiar with interaction design, personalization, and AI system evaluation. This group provided formative feedback on the system before deployment with community gardeners.

\textbf{Phase III.} We recruited university students involved in campus community gardening ($n=5$) to study how active gardeners perceive CultivAgents and envision future relationship-centered multi-agent gardening support systems. This specific population is chosen because their hybrid status as both gardening interested and tech-literate users allows them to identify friction points that a purely technical team might overlook.

\subsection{Data Collection}


\textbf{Phase I.} Experts interacted with CultivAgents and provided feedback on its cultural knowledge, ecological specificity, and responses to context-specific gardening use cases. This feedback informed refinements to the system's cultural and environmental grounding.

\textbf{Phase II.} HCI researchers completed a pre-survey, interacted with CultivAgents, and completed a post-survey. The surveys included 5-point Likert-scale items, along with open-ended questions. The pre-survey captured participants' gardening background, confidence, motivation, prior AI experience, and interest in culturally meaningful plants. The post-survey measured perceived usability, usefulness, trust, care, agent specialization, motivation, knowledge gain, and open-ended reflections on system strengths and limitations. The post-survey also included items adapted from the System Usability Scale (SUS)~\cite{brookeSUSQuickDirty1996}. Participants also provided formative feedback on usability, agent roles, interaction flow, and response quality. This feedback informed system refinements, including improvements to agent coordination, conversational memory, response formatting, and chat interface rendering.

\textbf{Phase III.} Community gardeners completed the same pre/post survey protocol used in Phase II. The workshop included three activities: system interaction, gap identification, and persona prototyping. First, participants were introduced to CultivAgents and used tablets to interact with the Experience, Environmental, and Ethnobotanical agents. Next, a facilitated discussion invited participants to identify gaps in the agents' knowledge, tone, and contextual support, such as location-specific growing constraints or the need for more tailored guidance for particular gardening challenges. Finally, participants created ``Persona Cards'' for new or revised specialized agents to recognize individuals’ sovereignty in their own gardening
journey. These activities allowed us to evaluate the current prototype while eliciting design directions for future relationship-centered multi-agent gardening systems.

\subsection{Data Analysis}

\textbf{Phase I.} We used expert feedback to identify gaps in CultivAgents' cultural knowledge, ecological specificity, and contextual relevance. 

\textbf{Phase II.} We analyzed survey responses  to examine HCI researchers' perceptions of usability, motivation, confidence, trust, care, knowledge gain, and the value of specialized agents. For SUS items, we calculated standard SUS scores and report descriptive statistics. We also reviewed open-ended responses to identify recurring usability issues and opportunities for system refinement.

\textbf{Phase III.} We used descriptive quantitative and thematic analyses. Quantitatively, we summarized pre/post survey responses to examine participants' gardening motivation, confidence, perceived knowledge gain, trust, care, usability, and perceptions of specialized agents, and compared pre/post responses descriptively when applicable. 

Qualitatively, we conducted a multi-phase \emph{thematic analysis} of open-ended survey responses, workshop discussion notes, and persona prototyping artifacts. This analysis focused on how participants described relationships with the system, plants, cultural knowledge, and community, as well as how they imagined future agent roles. We first reviewed the qualitative materials to identify recurring patterns, then established primary relational themes (Human-to-AI, Human-to-Plant, Human-to-Human, and AI-to-AI). Following theme stabilization, two researchers performed an axial coding pass to apply 12 specialized subcodes across 62 data units. To ensure analytical rigor, we calculated inter-rater reliability between the two coders, achieving 87.1\% raw agreement (54/62 units) and Cohen's $\kappa = 0.87$, indicating ``almost perfect'' agreement~\cite{Landis1977-pu}. Workshop artifacts were used to triangulate participants' reflections with their proposed persona designs, including desired agent expertise, tone, and forms of contextual support.

\section{Findings}
Our findings draw from three stages of evaluation: expert feedback with gardening and agricultural education collaborators, a formative workshop with HCI researchers, and a participatory workshop with community garden participants. Across these studies, we found that CultivAgents helped users move from general interest in gardening toward more situated, actionable knowledge. Also, participants and experts surfaced important limitations around cultural specificity, ecological grounding, and coordination across agents.
\begin{table}[t]
  \centering
  \small
  \setlength{\tabcolsep}{2pt}
  \renewcommand{\arraystretch}{1.08}
  \caption{Participant demographics across the formative HCI researcher workshop and community garden workshop. }
  \label{tab:demographics}
  \vspace{-1mm}
  \begin{tabular}{lccp{2.1cm}lll@{}}
    \toprule
    ID & Age & Gender & Cult. Background & Exp. & Cult. plants & Chatbot use \\
    \midrule
    \multicolumn{5}{l}{\emph{HCI researchers}} \\
      H1 & 18--24 & F & European & Inter. & Wants to & Occas. \\
      \multirow{2}{*}{H2} & \multirow{2}{*}{35--44} & \multirow{2}{*}{M}
      & Hawaiian /
      & \multirow{2}{*}{Exper.} & \multirow{2}{*}{Yes} & \multirow{2}{*}{Daily} \\
      & & & Pacific Islander & & & \\
      H3 & 25--34 & F & European & Novice & Wants to & Never \\
      H4 & 25--34 & NB & European & Novice & Wants to & Occas. \\
      \multirow{3}{*}{H5} & \multirow{3}{*}{35--44} & \multirow{3}{*}{F}
      & Hispanic /
      & \multirow{3}{*}{Novice} & \multirow{3}{*}{Wants to} & \multirow{3}{*}{Rarely} \\
      & & & Latino/a/x / & & & \\
      & & & Multiethnic & & & \\
      H6 & 25--34 & M & European & Novice & Wants to & Daily \\
      H7 & 25--34 & F & European & Inter. & Wants to & Regularly \\
    \addlinespace[2pt]
    \multicolumn{5}{l}{\emph{Community garden}} \\
      P1 & 25--34 & F & European & Inter. & Not sure & Rarely \\
      P2 & 18--24 & F & Asian & Exper. & Wants to & Regularly \\
      P3 & 18--24 & F & Asian & Inter. & Yes & Occas. \\
      P4 & 25--34 & F & European & Novice & Not sure & Never \\
      P5 & 18--24 & M & African-American & Beginner & Wants to & Regularly \\
    \bottomrule
  \end{tabular}
  \begin{minipage}{0.99\linewidth}
  \footnotesize
  \textit{Note.} ``Cult. Background'' refers to self-identified cultural background. Inter. = intermediate; Exper. = experienced; Occas. = occasionally; NB = non-binary / third gender. ``Cult. plants'' indicates whether participants reported growing, or wanting to grow, crops or plants that are culturally significant to them or their family. `Exp.' indicates gardening experience.
  \end{minipage}
  \vspace{-1mm}
\end{table}

\subsection{Expert Interviews}

Expert feedback suggested that while CultivAgents could support culturally relevant gardening reflection, its responses were often limited by generalized LLM knowledge and uneven digital representation of local agricultural practices. Experts noted that recommendations sometimes aligned with familiar cultural practices, but frequently remained surface-level or collapsed distinct cultural contexts into broader regional associations. For example, after one expert identified Ilokano, CultivAgents suggested ube and used Tagalog terminology such as ``gabi'' for taro rather than the Ilokano term ``aba.'' This response reflected a broad association with Filipino culture rather than attention to the user's specific linguistic and cultural context. These findings highlight how culturally oriented agents can unintentionally reproduce cultural flattening when finer-grained local knowledge, community validation, and mechanisms for clarification are absent.



The system also produced ecological grounding failures, including recommendations that conflicted with users’ described growing conditions or local ecosystems. Experts identified cases where suggested plants were poorly matched to wet soil conditions or included species inappropriate or invasive for the user's location (e.g., mangroves in Hawai`i). They also noted that practical constraints such as apartment restrictions and dense urban shade require more nuanced understanding than generalized climate data alone. These findings underscore the need for region-specific ecological knowledge, stronger conversational grounding, and community expert validation in culturally situated gardening agents.


\subsection{Formative Study with HCI Researchers}

Participants reported generally positive usability, with a mean SUS score of 73.93 and a median of 75.00, above the commonly used average SUS benchmark of 68~\cite{bangor2008empirical}. Gardening excitement remained stable from pre- to post-survey, with a mean of 3.71 before and after interaction, while mean gardening confidence increased from 2.57 to 3.14. After using CultivAgents, five of seven participants agreed that the system made them more motivated to start or continue gardening, and six of seven agreed that it gave them information they did not already know.

Participants responded positively to the specialized agent design. All seven participants agreed that the Environmental Agent gave location-relevant advice, and six of seven agreed that the Experience Agent understood their gardening skill level. Five of seven agreed the Ethnobotanical Agent provided culturally relevant information. Open-ended responses further suggest that participants valued the agents' distinct perspectives. H4 enjoyed having ``distinct agents'' that provided contextually appropriate answers through different lenses, while H2 appreciated the integration of ``multiple personas and contrasting viewpoints.'' H3 specifically valued the Cultural Agent, noting that its historical context and gardening information were most helpful. H6 also highlighted the system's ability to provide information at the intersection of environment and experience level.

However, the formative workshop also revealed limitations in the multi-agent experience. Although five participants agreed that multiple specialized agents were better than one general-purpose gardening chatbot, open-ended responses showed that the agents did not always feel coordinated. H6 noted that the agents did not seem aware of each other's responses, while H5 similarly felt that agents did not always speak to or build on one another. H3 and H5 noted that agents sometimes repeated similar information, and H2 observed that some agents appeared to lack full context or memory of the conversation. H1 was also confused by receiving multiple messages with different information. These responses suggest that participants valued agent specialization, but expected the agents to function as a coordinated system rather than as separate chatbots.

Participants also wanted stronger support for follow-up questions and conversational continuity. H7 wished they could ask the system to ``dig deeper'' into specific aspects of a recommendation, such as how to water basil appropriately, but found that the advice did not iterate much and sometimes repeated similar answers. H2 noted that some responses did not appear to retain memory of prior conversation context, and H5 wanted agents to build more directly on what the user had already asked. Based on this feedback, we refined CultivAgents before the community garden workshop by improving short-term conversational memory, reducing repeated responses, and strengthening agent coordination.

\subsection{Workshop with Community Garden}
\subsubsection{CultivAgents strengthened users' confidence, knowledge, and perceived ability to take action} 

Participants in the community garden workshop entered the study with high excitement about gardening, which remained unchanged from pre- to post-study with a mean of 4.20 out of 5. This stable score likely reflects the recruitment context: participants were already connected to a community garden and interested in food growing. Therefore, CultivAgents' primary value was to help already interested participants feel more informed, confident, and prepared to act.

Across participants, we observed positive shifts in confidence, motivation, and trust in AI-supported gardening advice. Gardening confidence increased from a mean of 3.00 to 3.60, motivation increased from 4.00 to 4.40, and trust in acting on AI gardening advice increased from 3.20 to 4.00. Participants also strongly agreed that CultivAgents provided new information, with a mean of 4.60. These patterns suggest that CultivAgents helped participants translate existing gardening interest into more actionable knowledge.

Qualitative responses further illustrate this shift. P1 noted that they ``enjoyed this AI agent more than I thought I would,'' while P2 described the system as ``super fun to use'' and said that it ``really prompted more questions about gardening.'' P3 similarly valued how the Cultural Agent ``expands my knowledge past just the question I asked.'' These responses suggest that CultivAgents not only answered participants' questions, but also encouraged further inquiry and helped users imagine additional gardening actions they could take.

\subsubsection{CultivAgents was usable and easy to navigate}

Participants reported that the CultivAgents interface was friendly, clear, and easy to understand, with well-formatted responses. CultivAgents received a strong mean SUS score of 84.50, well above the commonly used average SUS benchmark of 68~\cite{bangor2008empirical}, suggesting high perceived usability overall.

Open-ended responses help explain these ratings. P2 noted that ``the UI felt very friendly'' and that it was ``very easy to understand the advice with the way the responses were formatted.'' P4 similarly appreciated that the information was ``laid out clearly'' and that they could focus on the advice they were most interested in while still receiving relevant information. Participants also felt that the system answered their questions well: P2 wrote that it ``answered everything well,'' P4 said that ``nothing was answered poorly,'' and P5 noted that ``it answered everything quite well.''

At the same time, participants identified opportunities for more guided interaction. P4 suggested that the system could prompt users toward a ``next logical question.'' This comment suggests that although CultivAgents was perceived as usable, future versions could better scaffold multi-step gardening tasks by helping users understand what to ask next.

\subsubsection{Specialized personas were valued, especially for environmental personalization}

Participants responded positively to the specialized agent design. The Environmental Agent received the strongest ratings, particularly for location relevance, with a mean of 4.80, and climate and seasonal accuracy, with a mean of 4.40. It was also selected as the most helpful agent by three of the five participants. The Experience Agent was rated positively for tailoring advice to users' gardening skill levels, while the Ethnobotanical Agent was viewed as culturally relevant and respectful.

Participants also supported the multi-agent structure. They agreed that multiple specialized agents were better than a single general chatbot, with a mean of 4.60, and that the agents complemented each other, with a mean of 4.20. P5 explained this value directly, noting that they liked having ``multiple agents for different purposes'' and would feel ``less confident'' relying on one agent to explain everything. P3 similarly wanted to ``see multiple viewpoints'' in order to decide what was best for themselves.

Participants especially valued specialized agents when they provided situated information that a generic chatbot might overlook. P1 requested ``hyperlocal climate/soil information,'' while P3 wanted the system to explain soil nutritional content based on a description. These responses suggest that the value of specialized personas came not only from dividing information across different roles, but also from making gardening advice feel more personalized, local, and actionable.

\subsubsection{Human-AI Relationship: Trust Through Supportive Expertise}

Participants framed CultivAgents as a trusted source of supportive expertise. Post-use trust in acting on its advice was high (mean = 4.00), and open-ended responses emphasized practical, encouraging, and instructional support. P1 wanted agents to be ``direct,'' ``informative,'' ``culturally relevant,'' and ``to-the-point practical.'' P2 preferred a ``warm, casual, encouraging + supportive'' tone, similar to P5, while P4 wanted ``a coach'' that could identify areas for improvement while affirming progress.

These responses suggest that participants wanted a human-AI relationship grounded in supportive expertise rather than social companionship. They valued warmth, patience, and encouragement, but primarily in service of learning, action, and confidence-building.

\subsubsection{Human-Plant Relationship: CultivAgents helped users attend to plants as living, situated beings}

Participants' responses suggest that CultivAgents supported human-plant relationships by making plant care more situated, actionable, and environmentally attentive. Rather than offering generic care instructions, participants valued guidance that helped them understand plants' needs in specific growing contexts and translate that understanding into concrete care practices.

Participants described plant care as an ongoing process that changes as plants grow. P1 wanted ``diagnostics for pests and infections'' with treatment options, while P3 wanted guidance on ``how to take care of the plant as it grows,'' including identifying diseases and leaf discoloration. P2 emphasized concrete practices such as ``planting/pruning'' and ``propagating.'' P4 wanted to take a picture of their plant or gardening setup and have an agent ``tell me what to fix/change'' and ``point out what I've done wrong or right,'' suggesting a desire for feedback grounded in the visible condition of their own plants.

Participants also framed plant care as part of a broader ecological relationship. P1 emphasized ``sustainable harvest,'' while P2 wanted gardening advice to be ``environmentally conscious'' and ``safe for wildlife.'' Workshop artifacts and survey responses further showed that participants valued hyperlocal information, such as neighborhood-specific soil conditions, as a way to anchor plant-care decisions in their own ecological contexts. These responses suggest that CultivAgents helped move plants from digital abstractions to situated living beings embedded within local ecosystems.


\subsubsection{Human-Human Relationship: Culturally situated gardening support as everyday food autonomy}

CultivAgents showed promise for supporting human relationships around gardening by connecting advice to cultural knowledge, community practices, and intergenerational learning. 
Participants strongly valued cultural or community knowledge in gardening advice, with a mean of 4.60.
Participants consistently described gardening as a shared and relational activity grounded in learning from elders, neighbors, and community gardeners.
For example, P3 valued community gardens as spaces for knowledge sharing, noting that ``one of my favorite things about gardening in a community garden is learning from others.'' P1 emphasized intergenerational learning through ``learning from an elder. Rather than replacing human knowledge holders, participants viewed CultivAgents as most valuable when it reinforced and extended community-based learning and cultural connection. These findings suggest that relationship-centered gardening agents should support, rather than disrupt, the social and cultural relationships that sustain local food-growing practices.

\subsubsection{AI-AI Relationship: Users valued coordinated agent collaboration over parallel responses}

Participants' responses suggest that the value of CultivAgents depended not only on having multiple specialized agents, but also on how those agents related to one another. Participants appreciated the multi-agent format when agents provided complementary perspectives, built on each other's responses, and collectively produced a more complete answer. P1 valued ``having different agents to provide a more in-depth answer,'' while P3 appreciated being able to ``see multiple viewpoints'' on their question. These responses suggest that users valued agent outputs that felt cumulative rather than redundant.

Participants also imagined more explicit forms of agent collaboration. P1 suggested that agents could work together by ``compiling an answer that best answers a question,'' while P2 wanted agents to suggest follow-up prompts that would allow another agent to contribute more deeply from its area of expertise. P4 described this interactional rhythm as a ``complementary tone,'' where agents could provide shorter, interleaved responses, ``almost like someone pausing for add[ed] context.'' These comments suggest that participants wanted the agents to behave less like separate chatbots answering in parallel and more like a coordinated team.

However, participants also noted risks when agent relationships were not well coordinated. P4 felt that the Cultural Agent was ``more disconnected from the other two agents,'' suggesting that specialization alone was insufficient without integration. Participants also raised concerns about contradiction: P3 noted that multiple agents may help beginners, but that ``it's better to not have too many contradictions,'' and P5 suggested that agents should have separate duties to avoid repeating or contradicting one another. These findings suggest that effective AI--AI relationships require clear role boundaries, complementary expertise, and visible coordination across agents.

\subsubsection{Design opportunities for relationship-centered gardening agents}
Participants' feedback points to several opportunities for designing future relationship-centered gardening agents. First, users wanted more visual support and multimodal interaction. P4 requested ``visual aids,'' while P5 noted that ``visuals are important in gardening advice.'' In the post-survey, both P4 and P5 specifically requested ``pictures.'' P4 also wanted to take a picture of their gardening setup and have an agent identify what to fix, change, or affirm. 

Second, participants wanted stronger support for planning, timing, and longer-term care. P2 requested ``scheduling/planning when to plant when'' and later suggested features for planting timing and schedules. P3 also asked the system to ``generate a gardening schedule.'' These comments suggest that users wanted CultivAgents to move beyond one-time question answering toward ongoing planning support for planting, pruning, harvesting, and seasonal care.

Third, participants wanted more contextual and comprehensive information. P1 requested ``hyperlocal climate/soil information'' and ``more invasive plant information,'' along with requests for science and financial agents to complement the current set of three. 

Finally, participants wanted the system to better scaffold inquiry through follow-up questions and adaptive explanations. P1 imagined agents that answer a question with ``targeted follow-ups,'' while P4 wanted prompts toward the ``next logical question.'' 


\subsubsection{Future Personas}
\begin{figure*}
  \includegraphics[width=\linewidth]{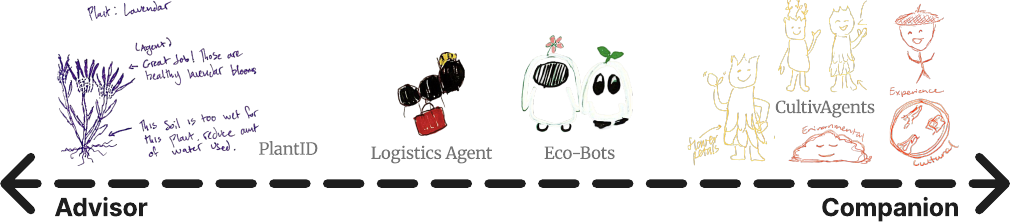}
  \vspace{-5mm}
  \caption{Participant-designed CultivAgents personas along an advisor--companion spectrum, 
  including companion-like naturalistic agents, an Eco-Bot, a Logistics Agent, and less anthropomorphized visual guidance.}
  \label{fig:persona}
  \vspace{-3mm}
\end{figure*}

Through co-design activities, participants extended CultivAgents by imagining new agent personas, visual forms, and interaction styles. As shown in Fig.~\ref{fig:persona}, P1 and P2 imagined friendly and gentle agents represented through naturalistic imagery, framing CultivAgents as companion-like supports. P3 designed an ``Eco-Bot'' that would provide sustainability- and science-focused metrics and insights. P5 designed a ``Logistics Agent'' represented as a worker ant to support garden planning, material costs, and sourcing. In contrast, P4 imagined a less anthropomorphized agent: a visual interaction that could provide technical guidance while affirming care practices.

 \vspace{-2mm}
\section{Discussion}
CultivAgents supported gardening as a situated, relational practice. Participants valued advice that connected local growing conditions, plant care, cultural knowledge, and complementary agent perspectives. They also surfaced limits in ecological grounding, cultural specificity, and agent coordination. We discuss these implications for relationship-centered multi-agent systems.

\subsection{From Advice to Situated Agency}
CultivAgents helped participants move from general interest to context-specific action. Gardening decisions depend on climate, soil, space, season, skill, plant variety, cultural context, and personal goals. By separating environmental, experiential, and cultural expertise, the system helped users interpret what advice meant for their own circumstances. This suggests that personalization should do more than adjust tone, difficulty, or recommendations. Gardening systems should help users reason through tradeoffs, understand why advice fits their context, and identify feasible next steps. In this sense, CultivAgents supported \emph{situated agency}: it did not replace gardeners' judgment, but helped them ask better questions and act with greater confidence.

\subsection{Coordinating Multi-Agent Expertise}

Participants valued multiple agents because gardening questions often combine ecological feasibility, skill-appropriate instruction, cultural meaning, cost, safety, and sustainability. A multi-agent structure can make these forms of expertise visible and approachable. However, specialization alone is not enough. When agents respond independently, the system can feel repetitive, fragmented, or contradictory. Future systems should therefore support coordinated expertise through shared memory, clearer role boundaries, contradiction checks, and synthesized responses. The goal is not to provide more voices, but to help users understand how different forms of expertise fit together.

\subsection{Relationship-Centered AI for Gardening}

A relationship-centered lens explains why participants valued CultivAgents beyond task completion. Gardening involves relationships among people, plants, places, and communities. Participants wanted agents that were warm but practical, attentive to plants as living things, grounded in local conditions, and respectful of cultural and community knowledge.

For human--AI relationships, this positions agents as trusted coaches or guides rather than necessarily companions. For human--plant relationships, systems should support longitudinal care through planning, reminders, visual feedback, and growth-stage awareness. For human--human relationships, systems should help to connect users to local gardeners, elders, extension services, seed libraries, and community organizations when knowledge is place-based or culturally specific. In multi-agent systems, relationship-centered design also applies to the agents themselves: how they share context, defer to one another, resolve disagreement, and present a coherent response.

\subsection{Cultural Grounding and Food Sovereignty}

Brown writes that ``the hardiest of plants and the most long-lasting solutions are those that emerge from indigenous soil, organisms that make connections and grow at their own pace. People do the same in their place-making.'' She framed place-making as something that grows through local relationships, histories, and conditions~\cite{Brown_2026}. CultivAgents raises a parallel design challenge: AI gardening systems should not only provide accurate plant-care advice, but also help users situate that advice within relationships to plants, place, culture, and community. Participants valued culturally meaningful guidance, yet expert feedback showed that cultural grounding remained uneven. The system produced stronger responses for well-documented agricultural regions, while advice related to Hawai`i was described as basic and, at times, culturally flattened. This limitation reflects broader concerns in Indigenous HCI that digital systems can extract or simplify cultural knowledge when they are not accountable to the communities from which that knowledge emerges~\cite{lewis2020indigenous}.

This unevenness points to a broader limitation: cultural grounding cannot be solved through prompting alone. If cultural advice depends on what knowledge is available, documented, and represented in AI systems, then designers of multi-agent gardening systems must make these limits visible rather than treating culture as a static user attribute. Instead of positioning AI as an authority on cultural practice, systems like CultivAgents should connect general guidance with local expertise, community knowledge, and lived experience. This reframes food sovereignty not as a system feature, but as a design orientation: supporting people's capacity to make informed growing decisions while remaining accountable to the places and communities from which agricultural knowledge emerges.

\subsection{Design Implications}

Our findings suggest four priorities for relationship-centered multi-agent systems in situated domains. First, systems should \textbf{support situated decision-making} by explaining why advice fits users' local conditions, resources, and goals. Second, they should \textbf{make collaboration visible} by showing how agents coordinate and synthesize expertise. Third, they should practice \textbf{cultural humility} by asking clarifying questions, avoiding broad identity assumptions, and inviting validation from users or local knowledge holders. Fourth, they should \textbf{scaffold ongoing care} by supporting observation, planning, reflection, and connection to local expertise.

\vspace{-3mm}
\section{Conclusion}
We presented \method, a relationship-centered multi-agent system for personalized gardening support. By coordinating experiential, environmental, and ethnobotanical perspectives, \method translates general gardening advice into guidance that is more responsive to users' skills, local conditions, and cultural contexts. Across expert feedback, HCI evaluation, and participatory evaluation with community gardeners, users valued the system's actionable, locally grounded, and culturally meaningful support. 


\bibliographystyle{ACM-Reference-Format}
\bibliography{references}

\appendix

\end{document}